\documentclass{article}
\usepackage{spconf}

\title{AlignDPO: Preference-Gated Alignment for Reducing Hallucination in Decoder-Only TTS}

\name{\shortstack{Xiao Zhou\thanks{This work was supported by ConnexAI.}, Oisín Turbitt,
      Kit Bower-Morris, Jonathan Carlton, \\
      Jamie Stacey, Kris Y. Hong}}
\address{ConnexAI, Manchester, UK \\
  \{xiao.zhou, oisin.turbitt, kit.bower-morris, jonathan.carlton\}@connex.ai, \\
  \{jamie.stacey, kris.hong\}@connex.ai}

\usepackage[T1]{fontenc}
\usepackage{textcomp}
\usepackage{amsmath,graphicx,url,booktabs,subcaption,amssymb}
\usepackage{algorithm, algorithmic}
\usepackage{enumitem}
\usepackage[hidelinks]{hyperref}
\usepackage{eso-pic}

\begin{document}

\AddToShipoutPictureBG*{%
  \AtPageLowerLeft{\raisebox{1.0cm}{\makebox[\paperwidth]{%
    \parbox{0.86\paperwidth}{\centering\scriptsize
      \copyright~2026 IEEE. Personal use of this material is permitted.
      Permission from IEEE must be obtained for all other uses, in any current or
      future media, including reprinting/republishing this material for advertising
      or promotional purposes, creating new collective works, for resale or
      redistribution to servers or lists, or reuse of any copyrighted component of
      this work in other works.}}}}}

\maketitle

\begin{abstract}
Decoder-only text-to-speech (TTS) models scale efficiently but remain prone to content hallucinations that arise from weak text--speech alignment during autoregressive generation.
We find that robustness is governed by a non-monotone relation to the sharpness of the alignment-bearing attention heads: a moderate degree is best, whereas over-sharpening is no better than the unaligned backbone and even less robust.
Guided by this, we present AlignDPO, a post-training method that reaches this moderate regime by folding a lightweight connectionist-temporal-classification (CTC) alignment term into Direct Preference Optimization (DPO), applied only to the chosen samples, with no architectural or inference-time change.
On the Seed-TTS-Eval English set, this significantly reduces the content-hallucination and word error rates relative to a strong DPO baseline and lowers the severe content-hallucination rate to $\sim$0.6\% (from 4.4\%); a listening study further finds it preferred for naturalness over both the backbone and that baseline.
Alignment is thus best learned and kept moderate rather than maximized or imposed at decoding.
Audio samples are available at \url{https://align-dpo-demo.vercel.app}.
\end{abstract}


\section{Introduction}
\label{sec:intro}

Recent advances in decoder-only \cite{Wang2023NeuralCL, Wang2025SparkTTSAE, Zhou2025IndexTTS2AB} and flow-based \cite{Chen2024F5TTSAF, Mehta2023MatchaTTSAF, park25b_interspeech} models enable high-quality zero-shot TTS. Despite their scalability and efficiency \cite{kwon2023efficient, tensorrt2025}, decoder-only models are prone to content hallucinations \cite{Wang2023NeuralCL, hussain2025koel}: omissions, repetitions, or fabrications of content words caused by weak text--speech alignment during autoregressive generation.
Unlike natural variation in fillers or function words (e.g., ``hmm'', ``a''), such content errors degrade semantic fidelity and are our focus.

Prior remedies introduce monotonic alignment through architectural changes, external aligners, or training-time attention guidance, while a complementary line constrains attention only at inference time: attention-constrained inference (ACI)~\cite{Wang2024AttentionConstrainedIF}, which we confirm helps an unaligned model. Preference optimization such as DPO~\cite{Rafailov2023DirectPO} improves quality but leaves alignment structure implicit.

This paper presents \textbf{AlignDPO}, a post-training method that refines this native alignment. Because the backbone is trained from scratch without Large Language Model (LLM) initialization, stable text--speech alignment emerges in a small set of attention heads \cite{Wang2024AttentionConstrainedIF}; AlignDPO strengthens them during preference optimization with a lightweight CTC term. The main contributions are:
\begin{itemize}[leftmargin=1.2em]
\item \textbf{A non-monotone sharpness--robustness relation.} The moderate sharpening reached during post-training gives the lowest hallucination and is robust to ACI, whereas over-sharpening is no better than the backbone and degrades sharply under the same inference-time constraint.
\item \textbf{One criterion, placed inside DPO and gated by preference.} A single length-normalized CTC score both \emph{identifies} the alignment heads and supervises them, so neither an external aligner nor a teacher is needed, and it is applied \emph{inside} DPO to the chosen samples only. At a moderate weight this beats a strong DPO baseline on hallucination (15.0\% to 11.3\%) and word error rate, cutting the severe rate to $\sim$0.6\%, while chosen-only gating preserves out-of-domain robustness.
\end{itemize}

Our experiments use a purely acoustic, single-codebook WavTokenizer~\cite{ji2024wavtokenizer} (40~tokens/s) without semantic supervision; we train a 530M backbone from scratch on 980~h of supervised data and post-train on 1,600~h of automatically-constructed preference data.

\section{Related Work}
\label{sec:related}

\noindent\textbf{Decoder-only TTS and robustness.}
Neural codec language models recast TTS as autoregressive token prediction over an LLM backbone~\cite{Wang2023NeuralCL,touvron2023llama}, and systems such as Spark-TTS~\cite{Wang2025SparkTTSAE} and IndexTTS2~\cite{Zhou2025IndexTTS2AB} reach strong zero-shot quality. Lacking an explicit text--speech alignment, their unconstrained attention makes them prone to \emph{content hallucinations} (skips, repetitions, run-ons) that degrade intelligibility.
Recent flow-matching alternatives~\cite{Mehta2023MatchaTTSAF,Chen2024F5TTSAF,park25b_interspeech} achieve high robustness but abandon our target decoder-only paradigm.

\noindent\textbf{Enforcing monotonic alignment.}
Prior work based on encoder--decoder architectures \cite{neekhara2024improving}, forced aligner \cite{Han2024VALLERR}, transducer-style modeling \cite{du2025vall, bataev2025tts}, inference-time constraints \cite{Wang2024AttentionConstrainedIF}, fixed teacher guidance \cite{Wang2025AttnGuidance}, repetition-aware decoding \cite{chen2024valle2}, or chain-of-thought prosody prediction \cite{xin2024ralle} often introduces intrusive design changes or explicit alignment dependencies, reducing deployment simplicity and modeling flexibility.
Closest in spirit, Attention-Constrained Inference (ACI) reshapes attention at decoding time~\cite{Wang2024AttentionConstrainedIF}, but acts externally on every step. These add architectural components, external aligners, teacher models, or decode-time machinery.

\noindent\textbf{Preference optimization in TTS.}
DPO~\cite{Rafailov2023DirectPO} and related schemes align TTS to preferences~\cite{zhang2024speechalign}, improving intelligibility, naturalness, and similarity~\cite{tian2024preference,zhang2025intp, hussain2025koel}, applying fine-grained, segment-level optimization~\cite{yao2025fine}, and extending via group-relative~\cite{sun2025f5rtts,liu2025grpotts} and structured AI feedback~\cite{yang2025rlaifspa} methods. Yet none adds an explicit \emph{alignment} constraint targeting hallucination, and some still rely on human annotation~\cite{yao2025fine}. AlignDPO instead places a preference-gated CTC term inside DPO, with automated preferences and one teacher-forced criterion to identify the heads. We isolate this mechanism by holding scale, data, and tokenization fixed across systems. ACI~\cite{Wang2024AttentionConstrainedIF} is the external baseline, sharing $\theta_{\text{base}}$ and enforcing alignment at inference rather than in training, while a strong automated DPO baseline isolates the contribution of the CTC term.

\section{AlignDPO}
\label{sec:method}

AlignDPO builds on a decoder-only backbone $\theta_{\text{base}}$, trained from scratch with optimized kernels such as FlashAttention~\cite{dao2022flashattention}. It switches to an eager attention implementation only to extract and regularize internal attention maps during alignment, leaving the backbone's pretraining and the deployed inference path on the optimized kernels.
We model the conditional distribution of acoustic tokens $\boldsymbol{y}$ given text $\boldsymbol{x}$ as $p_\theta(\boldsymbol{y}\mid\boldsymbol{x})$, optimized via a standard cross-entropy loss $L_{\text{CE}}$.
At inference, the model generates target tokens $\boldsymbol{y}_t$ conditioned on a prompt $(\boldsymbol{x}_p, \boldsymbol{y}_p)$ and target text $\boldsymbol{x}_t$, forming the sequence $[\boldsymbol{x}_p, \boldsymbol{x}_t, y_{\text{SOS}}, \boldsymbol{y}_p, \boldsymbol{y}_t, y_{\text{EOS}}]$, where $y_{\text{SOS}} / y_{\text{EOS}}$ are the start/end-of-sequence acoustic tokens.

The central design question is \emph{where} the monotonic-alignment constraint should be applied. AlignDPO applies it during preference optimization: the alignment-emerged heads of $\theta_{\text{base}}$ are identified with a single CTC criterion, and a preference-gated CTC term is folded into DPO, producing the model $\theta_{\text{DPO+M}}$.
For analysis we also consider alternative placements of the constraint: applying the same CTC uniformly in supervised refinement ($\theta_{\text{M}}$), DPO without the CTC term ($\theta_{\text{DPO}}$), and the inference-time alternative ACI~\cite{Wang2024AttentionConstrainedIF}.

\subsection{Alignment-Emerged Heads and Single-Criterion Identification}
\label{sec:heads}

In a from-scratch decoder-only backbone, a small number of attention heads emerge whose speech-to-text maps are near-monotonic, known as \emph{Alignment-Emerged Attention Maps} (AEAMs)~\cite{Wang2024AttentionConstrainedIF}. We locate them with a \emph{single}, training-consistent criterion: the length-normalized CTC score of each head's row-normalized speech-to-text map.

Following forward-sum alignment learning~\cite{Shih2021RADTTSPF,badlani2022one}, we append a blank column with prior $p_{\text{blank}} = 0.367$~\cite{Shih2021RADTTSPF} and renormalize each row, an effective blank of $0.268$, to obtain the augmented map $\tilde{\mathbf{A}}_{\text{sub}} \in \mathbb{R}^{T \times (L+1)}$ ($T$, $L$ the acoustic/text lengths; $\tilde{A}_{t,j}$ its entries), and score monotonicity by the length-normalized CTC negative log-likelihood,
\begin{equation}
\label{eq:ctc}
L_{\text{CTC}} = -\frac{1}{L}\log \sum_{\boldsymbol{\pi} \in \mathcal{B}^{-1}(\boldsymbol{z})} \prod_{t=1}^{T} \tilde{A}_{t,\pi_t},
\end{equation}
where $\mathcal{B}$ is the CTC collapse function and $\mathcal{B}^{-1}(\boldsymbol{z})$ is the set of monotonic paths for $\boldsymbol{z} = (1, \dots, L)$.
Lower $L_{\text{CTC}}$ means a sharper left-to-right mapping. We score every layer--head pair $(l,h)$ (denoted $L_lH_h$, the $h$-th head of the $l$-th layer) on $\theta_{\text{base}}$ from a single \emph{teacher-forced} forward pass over ground-truth text--speech pairs, and select $\mathcal{H} = \{ (l, h) \mid L_{\text{CTC}}(l, h) < \tau \}$ (we use $\tau{=}2$). Teacher forcing keeps identification cheap and uncorrupted. A single forward pass is far cheaper than autoregressive decoding, and feeding ground-truth tokens prevents the model from hallucinating during identification, which would otherwise corrupt the very attention maps being scored. The same $L_{\text{CTC}}$ doubles as the refinement objective below, so identification and supervision share one architecture-agnostic objective.

\noindent\textbf{Comparison to ACI.} Unlike ACI's Attention Sweeping~\cite{Wang2024AttentionConstrainedIF}, which combines an entropy cost with an alignment cost requiring an external forced aligner on the ground truth, our criterion is a \emph{single} CTC score that marginalizes over all monotonic paths, needs no external aligner, and is reused for refinement.

\noindent\textbf{Why not refine in pretraining?} Applying $L_{\text{CTC}}$ uniformly to all supervised data ($\theta_{\text{M}}{=}\arg\min_\theta (L_{\text{CE}}{+}\lambda L_{\text{CTC}})$ from $\theta_{\text{base}}$, with CTC weight $\lambda$) sharpens the AEAMs indiscriminately; it also forces eager attention to read the maps, forfeiting the optimized kernels that let pretraining learn pronunciation fast at scale. We instead gate the term by preference, reinforcing alignment only where the data is reliable, while pretraining keeps its high-performance kernels.

\subsection{Alignment Refinement via Preference-Gated DPO}
\label{sec:preference}

The alignment constraint is applied by folding a CTC term into DPO~\cite{Rafailov2023DirectPO}, training the model $\theta_{\text{DPO+M}}$ \emph{directly from $\theta_{\text{base}}$} (without the intermediate $\theta_{\text{M}}$ refinement). The optimization favors generations that are both perceptually preferred and more monotonically aligned.

\noindent\textbf{Preference data.} From a prompt pool $\mathcal{P}$ and a target-text pool $\mathcal{T}$ (normal and challenging texts), $\theta_{\text{base}}$ generates multiple randomized candidates $Y$ per item. Candidates are split by Word Error Rate (WER) into hallucination-free ($H^{+}$) and hallucinated ($H^{-}$) sets, and within each set a Pareto ranking~\cite{hussain2025koel} over \{WER, Speaker Similarity (SIM), Alignment\} selects a \emph{chosen} $S^{+}$ and \emph{rejected} $S^{-}$. A pair is retained only if $S^{+}$ also exceeds $S^{-}$ in SIM and both candidates agree across two independent ASR systems, avoiding marginal pairs. The pipeline is fully automated.
The Alignment ranking signal is produced by an external wav2vec2-CTC forced aligner~\cite{baevski2020wav2vec} on the generated audio, not by the model's own AEAMs; it supplies two scalars, the mean word-level posterior along the alignment path and a \emph{stretch ratio}, the fraction of phones lasting at least twice the mean phone duration of the prompt, which flags stalling. It differs from the Parakeet-CTC detector used for evaluation: head identification, preference construction, and the hallucination metric use three independent signals.

\noindent\textbf{Objective.} Each example is segmented as
\begin{equation}
[\,\underbrace{\boldsymbol{x}_p, \boldsymbol{x}_t, y_{\text{SOS}}, \boldsymbol{y}_p}_{\text{prompt}},\ \underbrace{\boldsymbol{y}_t, y_{\text{EOS}}}_{\text{completion}}\,],
\end{equation}
with DPO probabilities computed over the completion tokens. We optimize $\theta_{\text{DPO+M}}$ relative to the frozen reference $\theta_{\text{base}}$:
\begin{equation}
  L = L_{\text{DPO}}(\theta; \theta_{\text{base}}) + \lambda L_{\text{CTC}}(S^{+}; \mathcal{H}) \, .
\end{equation}
The CTC term is \emph{preference-gated}: it is applied to the \emph{chosen samples only}. The model thus reinforces monotonic alignment only on reliable generations, while the maps of rejected samples never enter the alignment gradient. Preference and alignment gradients then reinforce rather than interfere, unlike the uniform refinement $\theta_{\text{M}}$.

\section{Experimental Setup}
\label{sec:experiments}

\subsection{Evaluation Metrics}
\label{sec:metrics}

We report the following objective metrics throughout; subjective naturalness is assessed separately by a listening study.

\begin{itemize}[leftmargin=1.2em]

\item \textbf{SIM}: Speaker similarity computed from embeddings extracted by an ERes2Net speaker verification model.\footnote{\url{https://www.modelscope.cn/models/iic/speech_eres2net_sv_zh-cn_16k-common/}}

\item \textbf{pMOS}: Perceptual quality estimated using WhiSQA \cite{Close2025WhiSQANS},\footnote{\url{https://github.com/leto19/WhiSQA}} 
a Whisper-based non-intrusive model that predicts mean opinion scores directly 
from waveforms.

\item \textbf{WER} / \textbf{CER}: Word and character error rates computed using the Parakeet-TDT-0.6B-v2 ASR model.\footnote{\url{https://huggingface.co/nvidia/parakeet-tdt-0.6b-v2}}

\item \textbf{HAL}: Content hallucination rate (utterances with skipped, repeated, or fabricated words), from a forced-alignment detector over Parakeet-CTC-1.1B emissions. On a blind 160-utterance human-validated set ($\approx$53 per system), it has ${\sim}95\%$ recall at ${\sim}49\%$ precision (inter-annotator $\kappa{=}0.78$). It over-flags low-confidence rare words, but these shared false positives cancel under paired McNemar. The ranking also holds without the detector, so it does not hinge on per-system false-positive rates: \emph{human-labeled} hallucination counts decrease monotonically (\textbf{Base}/\textbf{DPO}/\textbf{DPO+M}: 24/10/5 of 53 clips each; $\chi^2$ for trend $p{<}0.001$), and WER agrees.

\item \textbf{SEV-HAL}: Severe content hallucination rate, a subset of HAL: the percentage of utterances with at least four hallucinated \emph{content} words (content vs.\ function words by a fixed closed-class list), with function words excluded.

\end{itemize}

\subsection{Base Model}
\label{sec:base_model}

Our experiments employ a 530M-parameter LLaMA-based \cite{touvron2023llama} decoder-only Transformer (14 layers, hidden size 1600, feed-forward 6400, 24 query and 6 key-value heads of dimension 64) in a VALL-E-style \cite{Wang2023NeuralCL} autoregressive formulation. This keeps the post-training modules applicable across modern AR codec-based systems.

Text is phonemized using \texttt{espeak-ng},\footnote{\url{https://github.com/espeak-ng/espeak-ng}} and speech is tokenized via \texttt{WavTokenizer} \cite{ji2024wavtokenizer} at 40~tokens/s (24~kHz audio).
The backbone (\textbf{Base}, $\theta_{\text{base}}$) is trained using standard cross-entropy loss $L_{\text{CE}}$ on 980~h of data derived from the GigaSpeech-M dataset \cite{GigaSpeech2021} via Whilter-filtering \cite{ravenscroft2025whilter}.

\subsection{Comparison Systems}
\label{sec:postalign_mitigation}

Our primary systems are listed below; we additionally study \textbf{DPO+M} variants that sweep the CTC weight $\lambda$, drop the preference gate (ungated), or initialize from the over-sharpened \textbf{Base+M}.
\begin{itemize}[leftmargin=1.2em]
\item \textbf{Base} ($\theta_{\text{base}}$): the supervised backbone.
\item \textbf{Base+M} ($\theta_{\text{M}}$): the CTC alignment term added during \emph{supervised} training of $\theta_{\text{base}}$ (the suffix ``+M'' denotes the added monotonic-alignment term), rather than inside DPO; it is the supervised-placement alternative in the where-to-apply comparison.
\item \textbf{DPO} ($\theta_{\text{DPO}}$): plain preference optimization from $\theta_{\text{base}}$.
\item \textbf{DPO+M} ($\theta_{\text{DPO+M}}$): our preference-gated alignment from $\theta_{\text{base}}$ (chosen-only CTC inside DPO).
\item \textbf{ACI}~\cite{Wang2024AttentionConstrainedIF}: a training-free, inference-time constraint, the inference-time counterpart to our training-time alignment.
\end{itemize}

\subsection{Training Configuration}

\textbf{Supervised training (Base, Base+M).} \textbf{Base} is trained from scratch under cross-entropy ($L_{\text{CE}}$) alone for up to 20~epochs of AdamW~\cite{Loshchilov2019AdamW} (learning rate $3{\times}10^{-4}$, cosine schedule); optimized kernels learn the text--speech mapping quickly at scale. \textbf{Base+M} initializes from \textbf{Base} and continues supervised training with the CTC alignment term added on the AEAM heads ($\lambda$ linearly annealed to~1 over the first 2000~steps, uniform over all utterances).

\textbf{Preference optimization ($\theta_{\text{DPO}}$, $\theta_{\text{DPO+M}}$).} The prompt pool $\mathcal{P}$ contains 600~h of YODAS \cite{Li2023YodasYD} speech (5--12\,s clips) filtered by inter-ASR agreement; 80\% undergo bandwidth extension to simulate enhanced real-world prompts. The target text pool $\mathcal{T}$ includes 500k normal sentences (5--40 words) and 3k LLM-generated challenging texts (3--139 words; rare words, phone numbers, tongue twisters) to stress-test alignment. Normal and challenging texts are paired with 1 and 10 prompts, respectively.
For each pair, 20--30 candidate completions are generated and partitioned into $H^{+}$ and $H^{-}$ by exact word match (one edit tolerated below ten words; any deletion forces $H^{-}$), then ranked by \{WER, SIM, Alignment\} (five scalars; non-dominated front per group, ties by normalized distance to the ideal point) to select $S^{+}$ and $S^{-}$; the final $\mathcal{D}_{\text{pref}}$ comprises 156k pairs (1{,}600~h). During this ranking, WER additionally uses a Metaphone-based fallback\footnote{Implemented with the \texttt{jellyfish} library: \url{https://www.jpt.sh/projects/jellyfish/}.} to tolerate homophone substitutions.
Both $\theta_{\text{DPO}}$ and $\theta_{\text{DPO+M}}$ are trained from $\theta_{\text{base}}$ for up to 10~epochs with AdamW (learning rate $6{\times}10^{-7}$, DPO KL penalty $\beta{=}0.05$, batch size~64); for \textbf{DPO+M}, $\lambda$ is annealed to~0.1 over the first 200~steps. Each system is selected by its own validation objective (DPO, or DPO${+}$CTC for \textbf{DPO+M}) with early stopping.

\subsection{Evaluation Protocol}

The test set is the English subset of Seed-TTS-Eval,\footnote{\url{https://github.com/BytedanceSpeech/seed-tts-eval}} from which we remove utterances overlapping our training data, leaving 706 sentences. We also use a separate hard set of 500 long utterances (up to 50 words: tongue twisters, spelled-out alphanumerics like postcodes and phone numbers, and disfluent conversational speech) to stress-test both preference gating and the inference-time ACI constraint, well beyond the short, clean Seed-TTS sentences.

Two questions organize the comparisons. \emph{(i)~Where should the constraint be applied?} We contrast supervised placement (Base+M) with placement inside DPO (DPO+M), and test initializing DPO from Base+M rather than from Base. \emph{(ii)~How much should it sharpen the heads?} We sweep the CTC weight, contrast gated with ungated application, and stress-test robustness with the inference-time ACI baseline.

\section{Results}
\label{sec:results}

\subsection{Native Alignment Heads}
\label{sec:res-heads}

\begin{figure}[!t]
\centering
\includegraphics[width=0.82\columnwidth]{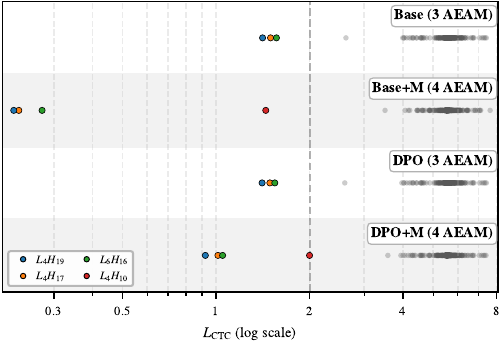}
\caption{Per-head $L_{\text{CTC}}$ distribution (log scale, teacher-forced over 100 utterances) for the four systems, sharing the x-axis. Colored dots are the identified AEAMs ($L_{\text{CTC}} < 2$); gray dots are the remaining heads.}
\label{fig:aeam_distribution_base_basem}

\vspace{2pt}
\includegraphics[width=0.82\columnwidth]{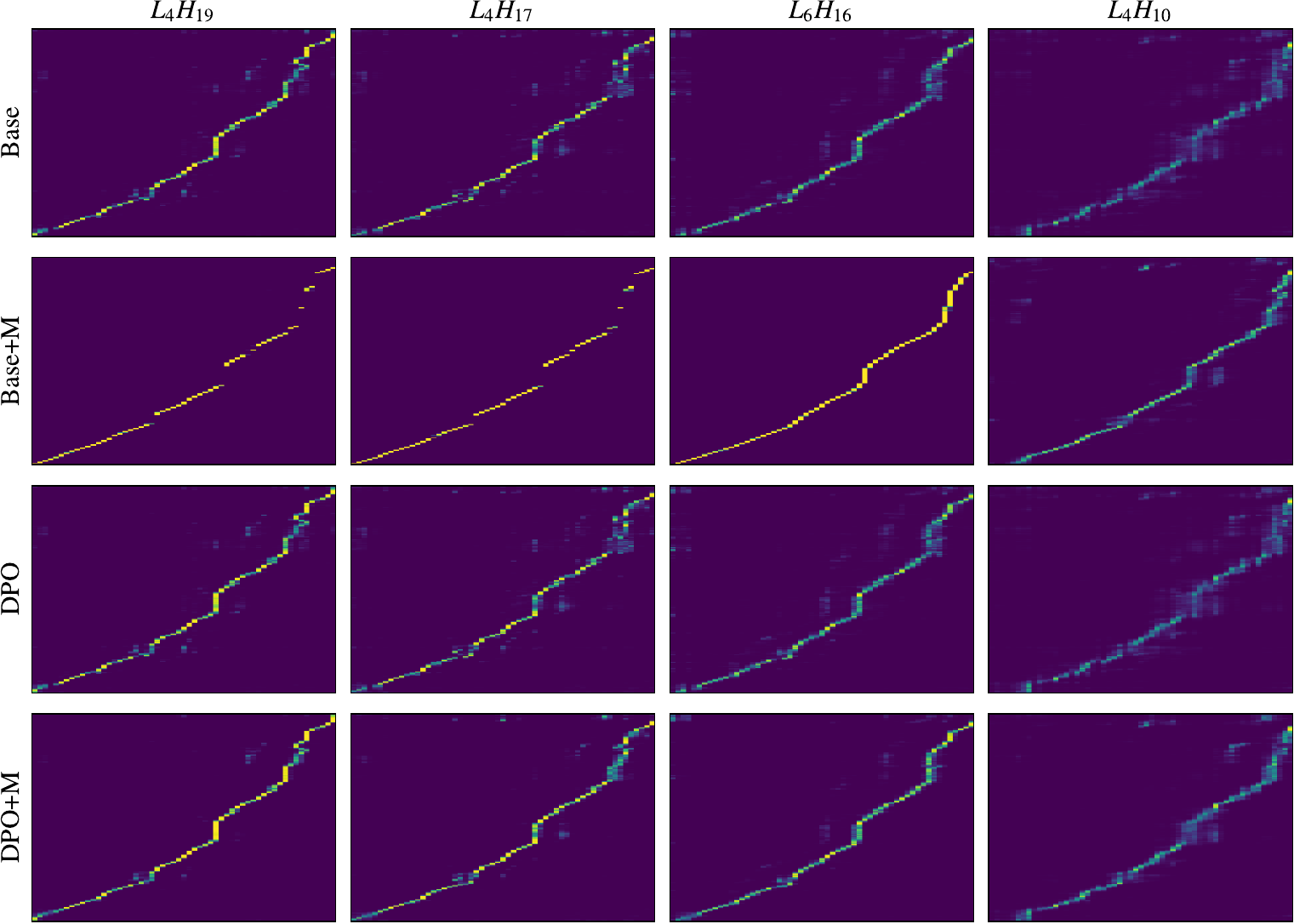}
\caption{Case study of the AEAM heads on one utterance across the four systems (rows), teacher-forced. Columns are the alignment-emerged heads $L_4H_{19}$, $L_4H_{17}$, $L_6H_{16}$, $L_4H_{10}$. Under teacher forcing \textbf{DPO} is nearly indistinguishable from \textbf{Base} (attention correlation ${>}0.999$); the maps reshape only with the CTC term ($+$M).}
\label{fig:aeam_case_study}
\end{figure}

As established by \cite{Wang2024AttentionConstrainedIF}, clear monotonic diagonals emerge in only a small subset of the \textbf{Base} model's heads, mostly in the lower and middle layers. Scoring every head by the CTC alignment loss $L_\text{CTC}$ identified exactly three such heads with $L_\text{CTC} < 2$ in \textbf{Base} ($L_4H_{19}$, $L_4H_{17}$, $L_6H_{16}$); \textbf{DPO} retained the same three (Fig.~\ref{fig:aeam_distribution_base_basem}).

Two phenomena were observed. First, although the CTC term is applied \emph{only} to Base's original three heads, the $+$M term pulled a fourth head, $L_4H_{10}$ ($L_{\text{CTC}}{\approx}2.9$ in Base), toward the alignment-emerged set: clearly below threshold for \textbf{Base+M} ($\approx1.7$) and to its edge for \textbf{DPO+M} ($\approx2.0$). The constraint thus not only sharpened the targeted heads but recruited an additional one. Second, as the case study in Fig.~\ref{fig:aeam_case_study} shows, \textbf{DPO+M} produced sharper, cleaner diagonals than \textbf{DPO}, reducing the residual diffuseness left by preference optimization alone (its per-head teacher-forced AEAM $L_\text{CTC}$ dropped from $\sim$1.4--1.5 to $\sim$1.0).

\begin{table}[t]
\centering
\caption{Objective metrics on Seed-TTS-Eval (English, 706 utterances). Higher SIM/pMOS and lower WER/CER/HAL/SEV-HAL are better. ``$+$ACI'' adds the training-free constraint~\cite{Wang2024AttentionConstrainedIF} to the row above; bold marks our system.}
\label{tab:main}
\small
\setlength{\tabcolsep}{3pt}
\renewcommand{\arraystretch}{0.88}
\begin{tabular}{lcccccc}
\toprule
\textbf{System} & \textbf{SIM}$\uparrow$ & \textbf{pMOS}$\uparrow$ & \textbf{WER}$\downarrow$ & \textbf{CER}$\downarrow$ & \textbf{HAL}$\downarrow$ & \textbf{SEV-HAL}$\downarrow$ \\
\midrule
Base            & 0.601 & 4.02 & 0.142 & 0.094 & 29.75 & 4.39 \\
\quad$+$ACI     & 0.601 & 4.01 & 0.113 & 0.071 & 22.24 & 2.41 \\
DPO             & 0.617 & 4.11 & 0.070 & 0.040 & 15.01 & 1.42 \\
\quad$+$ACI     & 0.614 & 4.11 & 0.069 & 0.037 & 14.87 & 1.42 \\
\textbf{DPO+M}  & \textbf{0.621} & \textbf{4.13} & \textbf{0.052} & \textbf{0.027} & \textbf{11.33} & \textbf{0.57} \\
\quad$+$ACI     & 0.622 & 4.13 & 0.049 & 0.024 & \phantom{0}9.63 & 0.14 \\
\bottomrule
\end{tabular}
\end{table}

\subsection{Hallucination and Intelligibility}
\label{sec:res-main}

Table~\ref{tab:main} reports the objective metrics; \textbf{DPO+M} was the best system that leaves decoding unchanged. It significantly outperformed the strong \textbf{DPO} baseline on hallucination (McNemar over 126 discordant utterances, 50 vs.\ 76, $p{=}0.026$) and WER (paired bootstrap $p{=}0.0008$), and reduced the severe rate to ${\sim}0.6\%$ (Wilson 95\% CI [0.22, 1.45]). SIM and pMOS were comparable across systems. Human labels independently confirmed this ordering (Sec.~\ref{sec:metrics}: 24/10/5). The primary \textbf{DPO+M} vs.\ \textbf{DPO} comparison (WER, HAL) survived Holm correction.

\subsection{Applying the Constraint}
\label{sec:res-ablation}

\begin{table}[tb]
  \centering
  \caption{Attention sharpness ($C_E$; lower $=$ sharper), free-running $L_{\text{CTC}}$, and HAL, by system and initialization (TF/FR $=$ teacher-forced/free-running, $n{=}706$; Init.\ ``--'' marks the from-scratch backbone). Rows are sorted by free-running $L_{\text{CTC}}$: HAL tracks $L_{\text{CTC}}$, not $C_E$ (cf.\ Figs.~\ref{fig:lambda_sweep},~\ref{fig:aci_ucurve}).}
  \label{tab:hardness}
  \small
  \setlength{\tabcolsep}{4pt}
  \renewcommand{\arraystretch}{0.88}
  \begin{tabular}{llcccc}
  \toprule
  \textbf{System} & \textbf{Init.} & \textbf{$C_E$ (TF)} & \textbf{$C_E$ (FR)} & \textbf{$L_{\text{CTC}}$ (FR)} & \textbf{HAL}$\downarrow$ \\
  \midrule
  \textbf{DPO+M}  & Base   & 0.64 & 0.48 & \textbf{1.13} & \textbf{11.33} \\
  DPO             & Base+M & 0.09 & 0.07 & 1.60 & 14.31 \\
  DPO+M           & Base+M & 0.09 & 0.07 & 1.61 & 15.58 \\
  DPO             & Base   & 0.93 & 0.84 & 1.70 & 15.01 \\
  Base+M          & Base   & 0.09 & 0.08 & 2.12 & 28.61 \\
  Base            & --     & 0.94 & 1.01 & 2.67 & 29.75 \\
  \bottomrule
  \end{tabular}
\end{table}

\begin{figure}[tb]
  \centering
  \includegraphics[width=0.82\columnwidth]{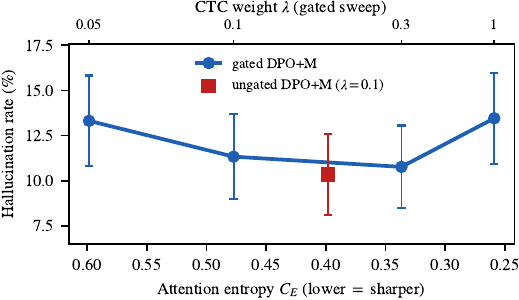}
  \caption{In-domain hallucination is non-monotone in attention sharpness (Seed-TTS, $n{=}706$). Each point is a separately trained \textbf{DPO+M} model, differing only in training config: the gated sweep varies the CTC weight $\lambda\in\{0.05,0.1,0.3,1.0\}$ on chosen samples; the ungated control adds rejected ($\lambda{=}0.1$). The $x$-axis is free-running $C_E$ (lower $=$ sharper); error bars are Wilson 95\% CIs.}
  \label{fig:lambda_sweep}
\end{figure}

We quantify alignment-head sharpness by the attention entropy $C_E = \tfrac{1}{T_s}\sum_t H(\mathbf{a}_t)$, the row entropy of the speech-to-text map averaged over the $T_s$ speech frames and the AEAM heads (natural log; lower $=$ sharper). Table~\ref{tab:hardness} reports it teacher-forced (TF) and free-running (FR); free generation is the sharper of the two once alignment is established, as the model's autoregressive commitments concentrate attention on a single text position, whereas the unaligned \textbf{Base} instead drifts and diffuses. The two systems initialized from \textbf{Base+M} isolated two design choices. First, refining alignment in pretraining is unnecessary: initializing DPO from \textbf{Base+M} rather than \textbf{Base} gave no significant difference in hallucination ($p{=}0.74$), and \textbf{Base+M} alone did not beat \textbf{Base} ($p{=}0.64$). Second, the CTC term is best applied \emph{once}, in DPO from raw \textbf{Base}: starting from the over-sharpened \textbf{Base+M} inherited its low $C_E$ but forfeited the DPO+M gain, reverting to the hallucination level of \textbf{DPO} (worse than the from-\textbf{Base} champion, $p{=}0.011$).

\subsection{Increasing Sharpness}
\label{sec:res-hardness}

Sharpness ($C_E$) is what $\lambda$ (in training) and ACI's radius (at inference) tune; hallucination, however, tracks \emph{misalignment} (the free-running $L_{\text{CTC}}$). Increasing $\lambda$ at fixed chosen-only gating reduced $C_E$ monotonically (Fig.~\ref{fig:lambda_sweep}). HAL, however, was non-monotone in sharpness: it was far higher at both the under-sharpened \textbf{Base} and the over-sharpened \textbf{Base+M} than in the moderate band ($\lambda{\approx}0.1$--$0.3$). Sharpness and alignment thus decoupled under heavy sharpening. The \textbf{Base+M}-initialized systems were far sharper than the moderate champion yet hallucinated well above it (Table~\ref{tab:hardness}). \textbf{Base+M} itself was the \emph{confidently misaligned} extreme: as sharp as any model, yet hallucinating almost as much as \textbf{Base}. HAL therefore tracks $L_{\text{CTC}}$, not $C_E$: a low $C_E$ is necessary, not sufficient.

As an ungated control, we applied the CTC term to chosen \emph{and} rejected samples ($\lambda{=}0.1$). In-domain, gating barely mattered: this control was statistically indistinguishable from gated \textbf{DPO+M} in HAL, if anything marginally lower (McNemar $p{=}0.57$, Table~\ref{tab:hard}). It became critical only on hard, out-of-domain inputs, where the term on rejected samples encoded spurious alignments, imperceptible on short utterances but evident on long ones. On the hard set, the ungated model regressed to the \textbf{DPO} baseline (matched WER, paired bootstrap $p{=}0.67$; matched severe rate), whereas gated \textbf{DPO+M} performed best (WER below the ungated control at $p{=}0.005$ and \textbf{DPO} at $p{=}0.019$; lowest severe rate). The binary detector saturated near $67\%$ out-of-domain and no longer separated the systems, so robustness was read from word error and severity. Gating therefore preserved out-of-domain robustness rather than reducing in-domain hallucination.

\begin{figure}[tb]
\centering
\includegraphics[width=0.82\columnwidth]{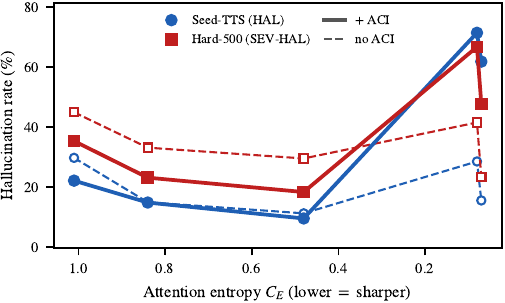}
\caption{ACI's effect is non-monotone (U-shaped) in attention sharpness, on both an in-domain and a hard out-of-domain set: hallucination vs.\ free-running entropy $C_E$ for five models. Left to right (soft to sharp): \textbf{Base} ($C_E{=}1.01$), \textbf{DPO} ($0.84$), \textbf{DPO+M} ($0.48$), \textbf{Base+M} ($0.08$), and \textbf{DPO+M} from \textbf{Base+M} ($0.07$). Seed-TTS reports HAL ($n{=}706$); Hard-500 reports SEV-HAL.}
\label{fig:aci_ucurve}
\end{figure}

Attention sharpness also governs decoding-time robustness. The training-free ACI constraint sets each head's window radius from its entropy, by ACI's own rule~\cite{Wang2024AttentionConstrainedIF} ($\rho{=}\mathrm{round}(8C_E){+}1$), applied to each model's own AEAM heads. Its effect was non-monotone (U-shaped) in $C_E$ on both test sets (Fig.~\ref{fig:aci_ucurve}): ACI helped the soft \textbf{Base} and was lowest at the moderate \textbf{DPO+M}, the best of any system on either set. On the over-sharpened models the entropy-derived radius collapsed to $\rho{=}2$ on the alignment heads, versus $5$--$10$ for the softer models. Accuracy then broke down: ACI drove both \textbf{Base+M} and \textbf{DPO+M} from \textbf{Base+M} ($C_E{\approx}0.07$) sharply upward, with the same collapse on the hard set. A control ruled out a narrow-window artifact: with a wide fixed radius ($\rho{=}7$, not the entropy-derived $\rho{=}2$), HAL for the over-sharpened \textbf{DPO+M} fell from $61.9\%$ back to $17.4\%$, matching plain decoding ($15.6\%$) but no better. The collapse was therefore intrinsic; no radius rescued an over-sharpened model. Both training (AlignDPO) and inference (ACI) thus showed that the moderate, well-aligned regime is best.

\begin{table}[tb]
\centering
\caption{Gating across regimes: in-domain WER and HAL (Seed-TTS, $n{=}706$) vs.\ hard-set WER and SEV-HAL (WER by paired bootstrap). The binary detector saturates ($\sim$67\%) out-of-domain and is omitted there.}
\label{tab:hard}
\small
\setlength{\tabcolsep}{5pt}
\renewcommand{\arraystretch}{0.88}
\begin{tabular}{lcccc}
\toprule
 & \multicolumn{2}{c}{\textbf{In-dom.}} & \multicolumn{2}{c}{\textbf{Hard set (OOD)}} \\
\cmidrule(lr){2-3}\cmidrule(lr){4-5}
\textbf{System} & \textbf{WER}$\downarrow$ & \textbf{HAL}$\downarrow$ & \textbf{WER}$\downarrow$ & \textbf{SEV-HAL}$\downarrow$ \\
\midrule
DPO                       & 0.070 & 15.01 & 0.297 & 33.2 \\
ungated ($\lambda{=}0.1$) & 0.050 & 10.34 & 0.302 & 32.2 \\
\textbf{DPO+M} (gated)    & 0.052 & 11.33 & \textbf{0.270} & \textbf{29.6} \\
\bottomrule
\end{tabular}
\end{table}

\subsection{Subjective Evaluation}
\label{sec:res-subjective}

HAL was measured objectively above; this study evaluates only whether reducing it affects \emph{naturalness}. We ran an A/B test (15 listeners, 30 utterances) on pairs manually verified to be hallucination-free across all systems, presented in randomized order with a ``no preference'' (N/P) option. Table~\ref{tab:abtest} reports the rates: by a per-listener Wilcoxon test, \textbf{DPO+M} was preferred over both \textbf{Base} ($p{=}0.001$) and \textbf{DPO} ($p{=}0.023$), and a two-sided binomial test on the decided votes agreed (over \textbf{Base} $p{<}0.001$, over \textbf{DPO} $p{=}0.004$). Our results demonstrate that suppressing hallucinations does not degrade naturalness but rather improves it.

\begin{table}[ht]
\centering
\caption{Subjective \emph{naturalness} preference (\%) over 15 listeners on utterances where both compared systems are hallucination-free. Bold marks the preferred system; N/P denotes ``no preference'' and $p$ is a two-sided binomial test on the decided votes; ``--'' marks the system absent from the pair.}
\label{tab:abtest}
\small
\setlength{\tabcolsep}{5pt}
\renewcommand{\arraystretch}{0.88}
\begin{tabular}{ccccc}
\toprule
\textbf{Base} & \textbf{DPO} & \textbf{DPO+M} & \textbf{N/P} & \textbf{$p$} \\
\midrule
28.2 & -- & \textbf{54.7} & 17.1 & $<$0.001 \\
-- & 33.6 & \textbf{46.0} & 20.4 & 0.004 \\
\bottomrule
\end{tabular}
\end{table}

\section{Conclusion}
\label{sec:conclusion}

\looseness=-1 AlignDPO mitigates content hallucinations in decoder-only TTS by folding a preference-gated CTC alignment term into DPO. Applied exclusively to chosen samples, it needs no architectural change, external aligner, or teacher. On Seed-TTS-Eval English it cuts the severe hallucination rate from 4.4\% to $\sim$0.6\% and improves word error and naturalness over a DPO baseline; out of domain, chosen-only gating preserves the gain that ungated training forfeits. This work uncovers a non-monotone sharpness--robustness relationship, showing that alignment is best kept moderate rather than maximized or imposed at decoding. We show this for one backbone, codec and corpus.

\noindent\textbf{Generative AI Use Disclosure.} Generative AI tools were used only for editing and polishing the manuscript's language. All technical content was developed independently by the authors, who remain fully accountable for the integrity of the work.

\end{document}